 \documentclass[prl,twocolumn,aps,superscriptaddress,showpacs]{revtex4}

\usepackage{graphicx}
\usepackage{color}
\usepackage{mathrsfs,amsmath,amssymb}

\begin{document}

\title{ Topological Insulator to Dirac Semimetal Transition Driven by
  Sign Change of Spin-Orbit Coupling in Thallium Nitride }

%\title{Thallium Nitride (TlN): A Novel Topological Insulator with opposite spin-orbit splitting}

\author{Xian-Lei Sheng}
\affiliation{Beijing National Laboratory for Condensed Matter
  Physics, and Institute of Physics, Chinese Academy of Sciences,
  Beijing 100190, China}

\author{Zhijun Wang}
\affiliation{Beijing National Laboratory for Condensed Matter
  Physics, and Institute of Physics, Chinese Academy of Sciences,
  Beijing 100190, China}

\author{Rui Yu}
\affiliation{International Center for Materials Nanoarchitectonics (WPI-MANA),
National Institute for Materials Science, Tsukuba 305-0044, Japan}

\author{Hongming Weng}
\email{hmweng@iphy.ac.cn}
\affiliation{Beijing National Laboratory for Condensed Matter
  Physics, and Institute of Physics, Chinese Academy of Sciences,
  Beijing 100190, China}
\affiliation{Collaborative Innovation Center of Quantum Matter, Beijing, China}

\author{Zhong Fang}
%\email{zfang@aphy.iphy.ac.cn}
\affiliation{Beijing National Laboratory for Condensed Matter
  Physics, and Institute of Physics, Chinese Academy of Sciences,
  Beijing 100190, China}
\affiliation{Collaborative Innovation Center of Quantum Matter, Beijing, China}

\author{Xi Dai}
%\email{daix@appy.iphy.ac.cn}
\affiliation{Beijing National Laboratory for Condensed Matter
  Physics, and Institute of Physics, Chinese Academy of Sciences,
  Beijing 100190, China}
\affiliation{Collaborative Innovation Center of Quantum Matter, Beijing, China}

\date{\today}

\begin{abstract}
  Based on the first-principles calculations, we reveal that TlN, a
  simple binary compound with Wurtzite structure, is a
  three-dimensional (3D) topological insulator (TI) with effectively
  negative spin-orbit coupling $\lambda_{eff} < 0$, which makes it
  distinguished from other TIs by showing opposite spin-momentum
  locking effect in its surface states. The sign of $\lambda_{eff}$
  depends on the hybridization between N-$2p$ and Tl-$5d$ states, and
  can be tuned from negative to postive by lattice strain or chemical
  substitution, which drive the system into a Dirac semimetal with 3D
  Dirac cones in its bulk states. Such topological phase transition can be
  realized by electronic mechanism without breaking any crystal
  symmetry.
%since TlN is Wurtzite instead of Zinc-blende structure. These intriguing quantum phenomena make TlN unique and a promising playground for quantum simulation in experiments.
\end{abstract}

\maketitle

%
%

%\section{Introduction}
%\section{introduction} \label{introduction}

\textit{Introduction}. The spin-orbit coupling (SOC) plays important
roles in generating topologically non-trivial band
structure.~\cite{Kane_PRL_2005,Bernevig_PRL_2006,Moore,Fu1,Fu2} For
examples, in graphene~\cite{Kane_graphene} (or other similar materials
like silicene~\cite{Yao_silicene}), the SOC opens a gap and makes it a
2D TI, while in Bi$_2$Se$_3$ family~\cite{Bi2Se3} 3D TIs, both the
band inversion and gap opening are caused by strong SOC. For the
design and optimization of topological electronic materials, it is
therefore highly desirable to have tunable SOC. From the viewpoint of
atomic physics, this goal is hardly achieved, because the strength of
SOC for each atomic orbital is almost predetermined by the type of
atoms (namely the atomic number). In a solid state compound, however,
we are interested in the effective SOC of certain Bloch states, which
in general compose of multiple atomic orbitals with different SOC.
This complication, on the other hand, gives us a chance to tune the
SOC effectively. Not only the strength but also the sign of SOC can be
effectively tuned through the proper manipulating of the orbital characters for
 low energy Bloch states, which
generates interesting topological phase transition as we will
addressed in the paper.

It has been indicated in literatures that several rare compounds
~\cite{Vidal,HgS,wangHgTe} may have effectively negative SOC. For
example, in HgS, it was suggested that the SOC splitting for the
$p$-orbitals around the valence band top (i.e, the $\Gamma$ point) is
opposite to that of the usual semiconductors with Zinc-blende
structure. In other words, the $j$=1/2 doublet states is energetically
higher than the $j$=3/2 quartet states in HgS, resulting in a TI state
rather than the zero-gap semimetal state in HgSe and HgTe. This
proposal is interesting, unfortuanately, the existence of band
inversion in HgS is challenged by the recent GW
calculations~\cite{gwHgS} and is not confirmed
experimentally~\cite{HgS2}. In this paper, we will demonstrate that
Wurtzite TlN~\cite{TlNfilm} is a promising TI with ``negative" SOC.
What makes it unique from other TIs is its opposite spin-momentum
locking effect, which can be readily observed experimentally.
Furthermore, the sign of effective SOC in TlN can be reversed by
suitable lattice strain without breaking any crystal symmetry, which
leads a topological phase transition from a TI to a Dirac semimetal
with 3D Dirac cones in its bulk.~\cite{BiO2,Na3Bi,Cd3As2} All of these
make TlN a valuable playground for quantum manipulation.

\textit{Mechanism for ``negative'' SOC.} In Zinc-blende III-V or II-VI
semiconductors, such as GaAs or CdTe, the conduction band minimum at
$\Gamma$ labeled as $\Gamma_{6}$ is mostly from cation $s$-orbital,
while the valence band maximum is mostly from anion $p$-orbitals,
which split into $\Gamma_{8}$ ($j=\frac{3}{2}$) and $\Gamma_{7}$
($j=\frac{1}{2}$) manifolds in the presence of SOC. The band gap $E_g$
is defined as the energy difference between $\Gamma_6$ and $\Gamma_8$
states, $E_g = E_{\Gamma_{6}}-E_{\Gamma_{8}}$, and similarly, the
effective SOC is defined as $\lambda_{eff} =
E_{\Gamma_{8}}-E_{\Gamma_{7}}$. Both $E_g$ and $\lambda_{eff}$ are
positive for GaAs and CdTe, while it is known that $E_g$ is negative
in HgTe (i.e, the $s$-like band is lower than the $p$-like bands)
leading to a topologically non-trivial state. Unfortunately, this
is a zero gap semimetal state (rather than a true insulator) due to the
four fold degeneracy of $\Gamma_{8}$ protected by the cubic symmetry
of the lattice. On the other hand, if both $E_g$ and $\lambda_{eff}$
are negative as suggested for HgS, the $\Gamma_7$ states, which are
only two-fold degenerated, will be higher than the $\Gamma_8$ states,
and the band inversion between the $\Gamma_6$ and $\Gamma_7$ states
will lead to a TI. Simultaneous requirements of both $E_g<0$ and
$\lambda_{eff}<0$ therefore make the possible material realization
difficult.

\begin{figure}[tbp]
\centerline{\includegraphics[clip,scale=0.40]{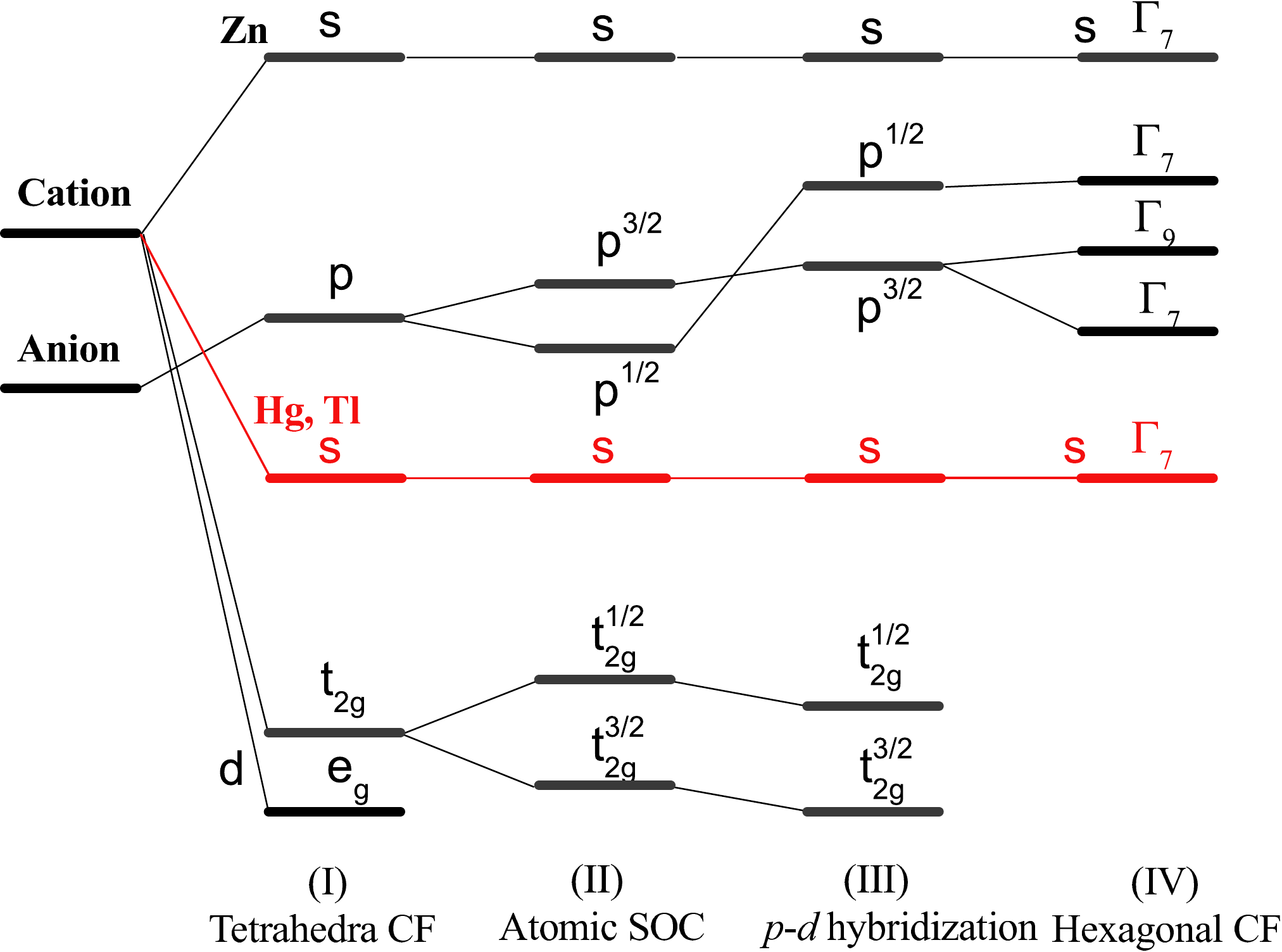}}
\caption{(Color online) The energy level diagram of Wurtzite semiconductor at $\Gamma$ by taking into account of (I) idea tetrahedral crystal field (CF), (II) atomic
spin-orbit coupling (SOC), (III) $p$-$d$ hybridization and (IV) hexagonally distorted CF in real structure. The $s$ orbital of Zn in ZnO is higher than anion $p$-orbitals,
while that of Hg in HgTe and HgS or Tl in TlN might be lower and result in inverted band structure.}
\label{bandinversion}
\end{figure}

To get $\lambda_{eff}<0$, a mechanism is illustrated as the following,
by takeing the Wurtzite ZnO as a example.~\cite{zno1, zno2} ZnO is a
normal semiconductor with $E_g>0$, however, its valence band top shows
negative SOC ($\lambda_{eff}<0$), mostly due to the presence of
shallow core $3d$ states of Zn. In ZnO, the Zn atoms are under
approximately tetrahedra crystal field of O, which leads to sizable
$t_{2g}$ and $e_g$ splitting of the Zn-$3d$ orbitals (with the
$t_{2g}$ states energetically higher than the $e_g$ states). We ignore
the $e_g$ states, and concentrated on the $t_{2g}$ manifold, which
mainly consists of three atomic orbitals, $d_{xy}$, $d_{yz}$ and
$d_{zx}$ orbitals. Considering the SOC effect among the three $t_{2g}$
orbitals (i.e, projection of SOC matrix into the subspace spanned by
these three orbitals), it is interesting to note that the SOC is
effectively negative.~\cite{SOCform} As the results, the $t_{2g}$ states
will further split into a $j_{eff}$=1/2 doublet and a $j_{eff}$=3/2
quartet states with the former energetically higher. This is in
opposite to the SOC splitting among the $p$ orbitals. If the $t_{2g}$
are shallow enough, they will hybridize with the O-$2p$ states
significantly, and the valence band top should be composed of both
$2p$ and $3d$ characters. Since the SOC among the O-$2p$ and
Zn-$t_{2g}$ atomic orbitals have opposite sign, the final effective
SOC of the valence band top therefore must be determined by its
relative composition, and can be tuned by controlling the $p-d$
hybridization in the system. For ZnO, in particular, the positive SOC
splitting among O-$2p$ orbitals are relatively smaller, the large $3d$
components of the valence band top leads to the effectively negative
SOC (Fig.\ref{bandinversion}).

%The reason is the following. We can map the $t_{2g}$ subspace into a space with %effective angular momentum $|L|$=1 by doing the mapping,

\begin{figure}[tbp]
\centerline{\includegraphics[clip,scale=0.40]{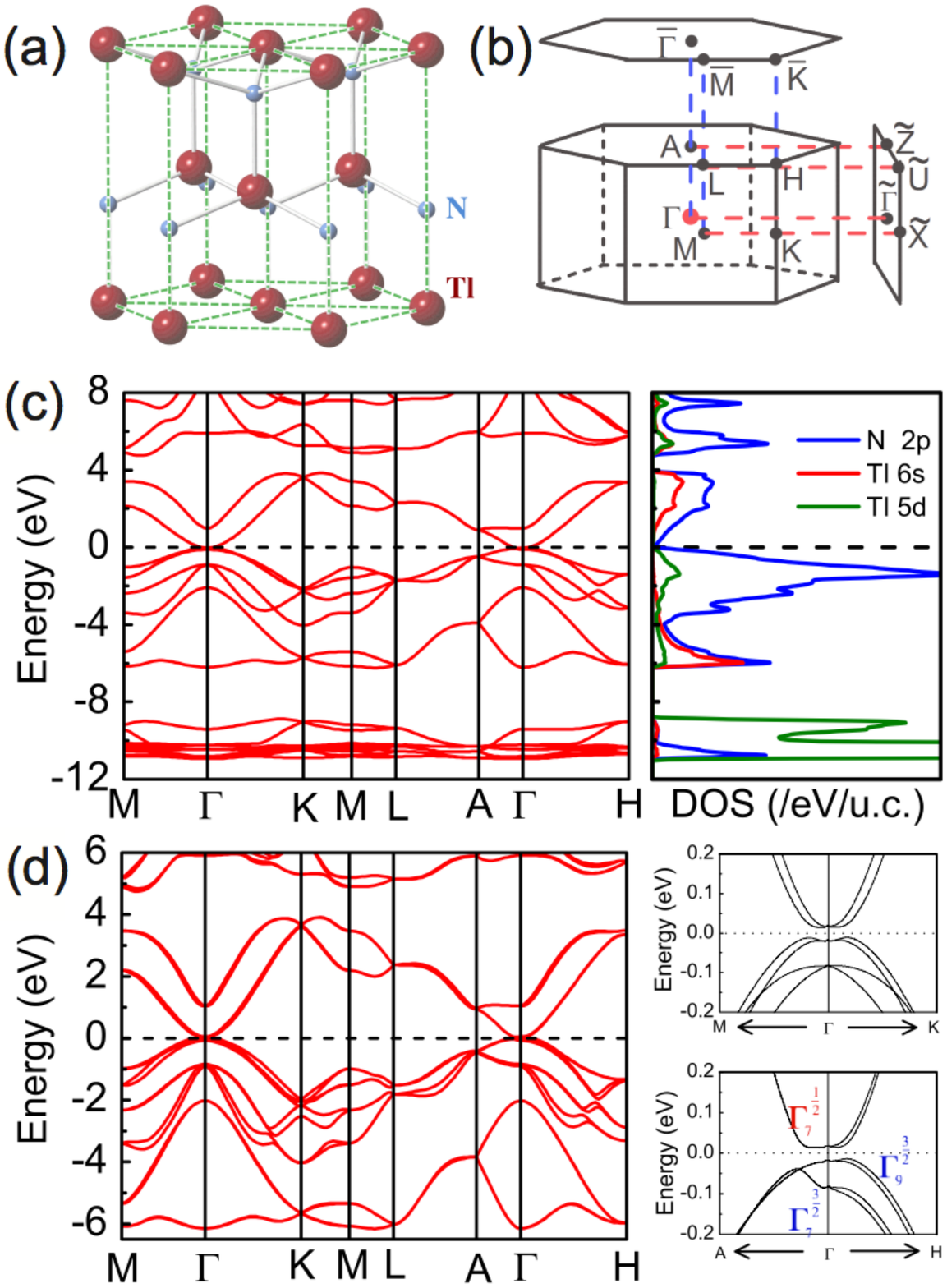}}
  \caption{(Color online) (a) Crystal structures of Wurtzite TlN and (b) its Brillouin zone (BZ), as well as the projected surface BZ of (0001) and (0100) surfaces.(c) The band structure and projected density of states of Wurtzite TlN without SOC. (d) The band structure with SOC and partially enlarged view around $\Gamma$ point.}
\label{band}
\end{figure}

\textit{Crystal structure and Methods.} TlN was synthesized in 1974
and it has Wurtzite structure~\cite{TlNfilm} (as shown in
Fig.\ref{band}(a)) with experimental lattice parameters
$a$=3.68~\AA ~and $c$=6.01~\AA. The Wyckoff position of N is (1/3,
2/3, 0) and that of Tl is (1/3, 2/3, 0.381). Wurtzite structure has
the $\cdots$AB-AB-AB$\cdots$ stacking sequence along (0001) direction,
and each primitive unit cell contains two formula units. Both Tl and N
are tetrahedrally coordinated with slightly hexagonal distortion. Our
first-principles calculations has been performed by using the WIEN2K
package~\cite{wien2k} with the full potential linearized augmented
plane wave method (FLAPW). The generalized gradient approximation
(GGA) is used for the exchange correlation potential. Brillouin zone
(BZ) integration with a mesh of 13$\times$13$\times$7 sampling is
used. The muffin-tin radii ($R_{MT}$) are chosen as 2.11 bohr both for
Tl and N. The maximum size of the plane wave vector ($K_{max}$) is
determined by $K_{max}R_{MT}=7.0$. For the later analysis, we
construct the maximally localized Wannier functions
(MLWF)~\cite{MLWF1997} for Tl $s$ and N $p$ orbitals, which
reproduce the low energy bands nicely. The MLWF was then used for the
effective Hamiltonian of both bulk and semi-infinite surface.

\textit{Electronic structure.} The band structures of TlN are
calculated without and with SOC included, respectively, as shown in
Fig.\ref{band} (c) and (d). The bands from -6 to 4 eV are mostly coming from the
N-$2p$ and Tl-$6s$ states. From the partial density of states (DOS),
it is easy to find that Tl-$6s$ state around the $\Gamma$ point
($\Gamma_{7}^{s}$) is lower than N-$2p$ by about 2.0 eV for both 
cases of with and without SOC. This suggests that TlN is a material
with band inversion (i.e., $E_g<0$), similar to the case of HgTe.
Considering the possible underestimation of band gap by the GGA, we
further check the band inversion by using modified Becke-Jonhson
potential~\cite{mBJ} and HSE method~\cite{HSE}, respectively, and
still find band inversion of about 1.5 eV, which is very large
compared to most known TIs. Thus, we conclude that the band inversion
in TlN is robust and is not an artifact of GGA type calculation.

The Tl $5d$-orbitals are fully occupied and locate mainly around -10.0
eV below the Fermi level. They are relatively extended with a sizable
band width of about 2.5 eV. There are apparently quite much $5d$ DOS
contribution to the N-$2p$ bands around the Femi level, which
indicates remarkable $p$-$d$ hybridization. This situation is quite
similar to that in ZnO and leads to effective ``negative" SOC for the
states around the valence band top. Due to the reduced symmetry (the
slightly hexagonal distortion), the $\Gamma_{8}$ ($j$=$\frac{3}{2}$)
states should further split into $\Gamma_{9}^{\frac{3}{2}}$ and
$\Gamma_{7}^{\frac{3}{2}}$ manifolds. Finally, as shown in
Fig.\ref{band}(d), the valence $2p$ band top in TlN split into three
groups: $\Gamma_{9}^{\frac{3}{2}}$, $\Gamma_{7}^{\frac{3}{2}}$ and
split-off band $\Gamma_{7}^{\frac{1}{2}}$. We find that their energies
descend in the order of $\Gamma_{7}^{\frac{1}{2}}$,
$\Gamma_{9}^{\frac{3}{2}}$ and $\Gamma_{7}^{\frac{3}{2}}$, which
sugests $\lambda_{eff}<0$. Thus, TlN satisfies both the conditions of
$E_g<0$ and $\lambda_{eff}<0$, and should be a 3D topological
insulator with negative SOC. Our calculations indeed found that it is
a semiconductor with the direct (indirect) band gap of 36meV (25meV)
and 25 meV. Since TlN has no inversion symmetry, the Wilson loop
method is employed to study the evolution of hybrid one-dimensional
Wannier function center~\cite{YuruiZ2, tireviewofus}, and its
$\mathbb{Z}_2$ topological indices are found to be (1;000) (see Supplemental Material for details).

\begin{figure}[tbp]
%\label{surface}
\includegraphics[clip,scale=0.4,angle=0]{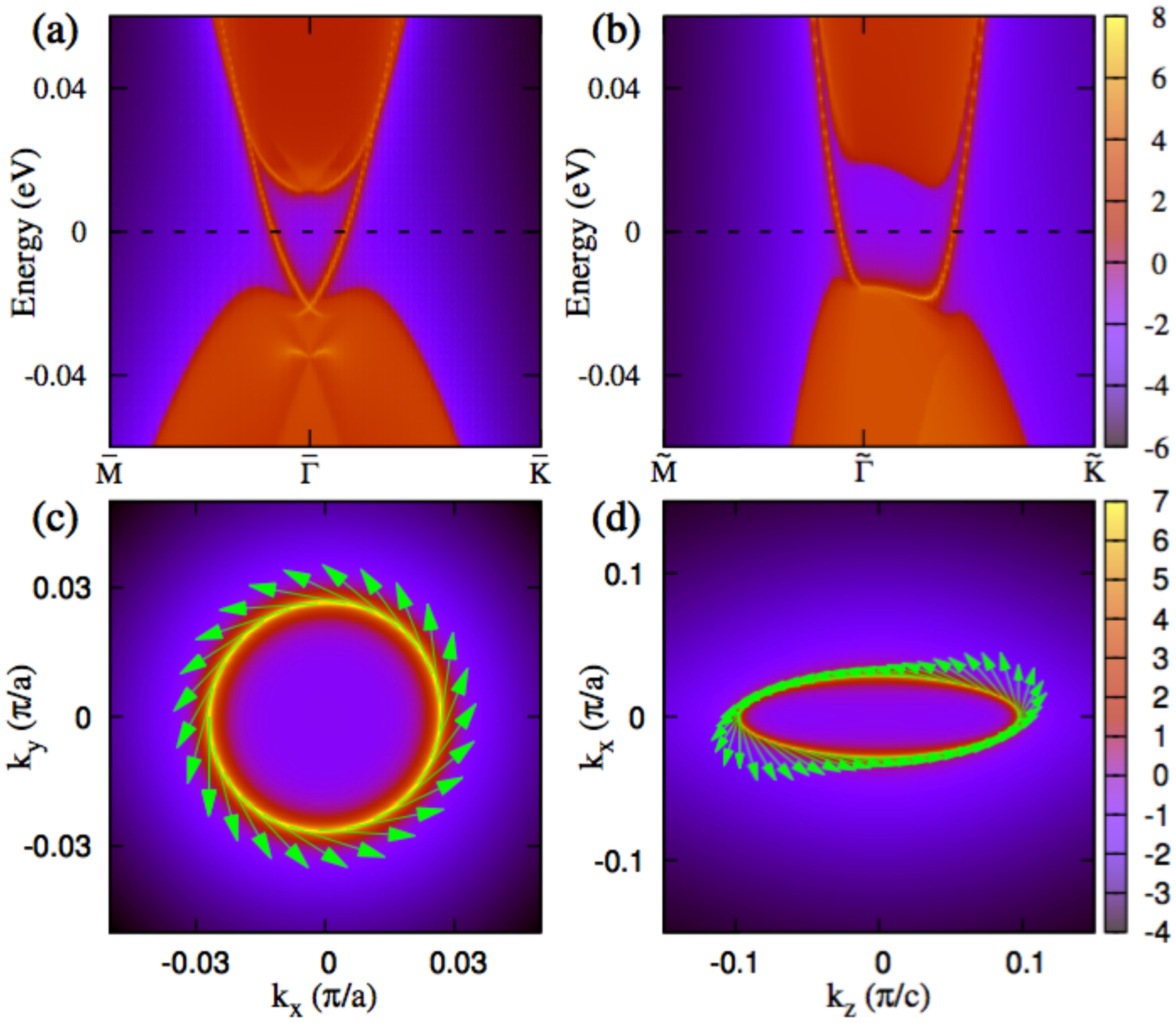}
%\centerline{\includegraphics[width=8.0cm,angle=270]{fig4.eps}}
\caption{(Color online)The calculated surface states (upper panels)
  and corresponding Fermi surfaces and spin textures (lower panels) of Wurtzite TlN on (0001) (left panels) and (0100) (right panels) surface, respectively.}
  \label{surface}
\end{figure}

3D strong TI supports odd number of Dirac-cone-like surface states,
which should have characteristic spin-momentum locking effect. Almost
all of the presently known TIs have shown left-handed spin-momentum
locking when Fermi level is above the Dirac
point.~\cite{Bi2Se3,T03Ag2Te,ZHJSOT,tireviewofus} As shown in
Fig.\ref{surface}, TlN does have single Dirac-cone-like surface states
around $\bar{\Gamma}$ for both (0001) and (0100) surfaces, while the
spin-momentum locking has {\it right-handed} helicity, in opposite to
all other known TIs up to now. It would be very interesting to check
this by spin-resolved ARPES experiments. It is also fundamentally
interesting to see what will happen in the interface formed by two TIs
with opposite spin-momentum locking helicity.~\cite{wangHgTe} Another
plausible remarkable thing is when $s$-wave superconducting pairing
interaction is introduced to them by proximity effect,~\cite{p+ip} the
induced topological superconducting state will have opposite
chirality, being $p$+$ip$ for left-handed one and $p$-$ip$ for
right-handed one. Heterostructure of them might have more fascinating
quantum phenomena.

\begin{figure}[tbp]
\includegraphics[clip,scale=0.35,angle=0]{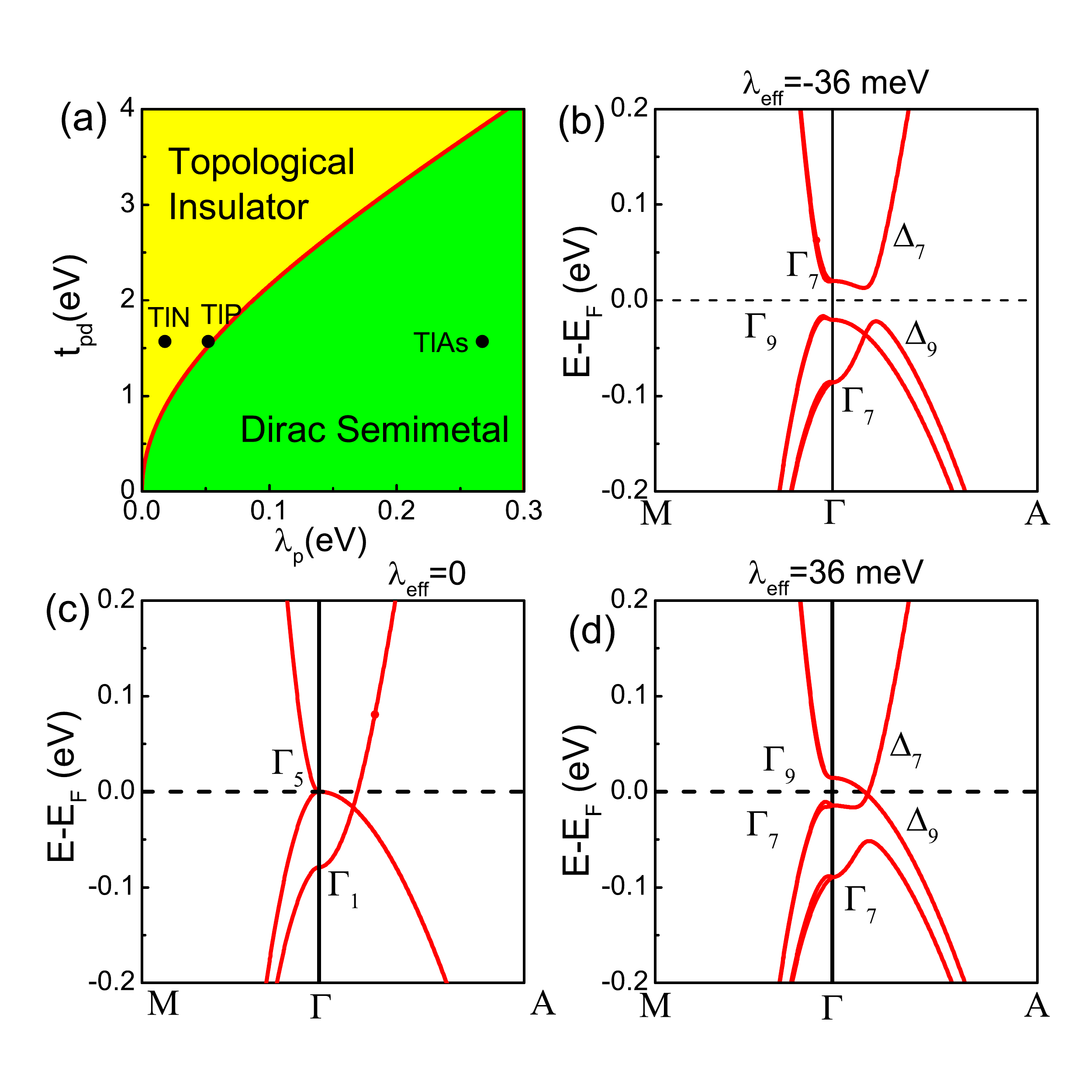}
\caption{(Color online) (a) Phase diagram of topological insulator and topological 3D Dirac semimetal depending on $p$-$d$ hybridization ($t_{pd}$) and atomic SOC of $p$ orbital ($\lambda_p$). The band structure with (b) $\lambda_{eff}$=-36 meV,
  (c) $\lambda_{eff}$=0 meV and (d) $\lambda_{eff}$ = -36 meV.}
\label{phase}
\end{figure}

\textit{$p$-$d$ hybridization}. As mentioned above, the $p$-$d$ hybridization~\cite{pdhybri} is the origin of ``negative" SOC splitting
in $p$ orbitals. To explicitly reveal the underlying physics, an effective Hamiltonian with the basis of N $2p$ and Tl
$5d$-orbitals is established. Here the Tl-$s$ orbital and the crystal field due to hexagonal distortion are neglected since
they have negligible effect on the effective SOC splitting in $p$ orbitals. Due to the tetrahedra crystal field, only $t_{2g}$
orbitals can have the hybridization with $p$ orbitals and $e_g$ part is ignored. For both $p$ and $t_{2g}$ orbitals, their
atomic SOC is firstly considered in their own subspace. The spin-orbit splitting of $t_{2g}$ is found to be opposite to
that of $p$ states with $D_{\frac{1}{2}}$ ($j$=$\frac{1}{2}$) higher than $D_{\frac{3}{2}}$ ($j$=$\frac{3}{2}$).

Based on this, we can further include the $p$-$d$ hybridization, parameterized as $t_{pd}$, to see its influence on the
effective spin-orbit spitting in $p$ orbitals. Only the $P$ and $D$ basis with the same representation can have nonzero
hybridization, i.e., $\langle P_{j}, j_z |H|D_{j'} ,  j_{z}' \rangle=t_{pd} \delta_{j, j'} \delta_{j_z , j_{z}'}$. Such a simplified model, in the basis set of $|P_{j}, j_z\rangle$ and $|D_{j}, j_z\rangle$ (with $j=\frac{3}{2},\frac{1}{2}$ and $j_z=-j,-j+1,\cdots,j-1,j$), can be written in second quantized form,
%We order thebasis with positive $j_z$ quantum number as following, $|P_{\frac{3}{2}},\frac{3}{2}\rangle$, %$|D_{\frac{3}{2}},\frac{3}{2}\rangle$,
%$|P_{\frac{3}{2}},\frac{1}{2}\rangle$, $|D_{\frac{3}{2}},\frac{1}{2}\rangle$, $|P_{\frac{1}{2}}, \frac{1}{2}\rangle$,
%$|D_{\frac{1}{2}},\frac{1}{2}\rangle$ and those with negative $j_z$ are followed. Therefore, the Hamiltonian reads
\begin{eqnarray*}
\label{eq:Hpd} H_{pd} = \sum_{j,j_z}E_{jj_z}^{p}\hat{c}^{\dagger}_{j,j_z}\hat{c}_{j,j_z}
+\sum_{j,j_z}E_{jj_z}^{d}\hat{d}^{\dagger}_{j,j_z}\hat{d}_{j,j_z}\\
 +\sum_{j,j_z}[t_{pd}\hat{c}^{\dagger}_{j,j_z} \hat{d}_{j,j_z} +t_{pd}^{*}\hat{d}^{\dagger}_{j,j_z}\hat{c}_{j,j_z}]
\end{eqnarray*}
where $\hat{c}_{j,j_z}$ ($\hat{d}_{j,j_z}$) and $\hat{c}^{\dagger}_{j,j_z}$ ($\hat{d}^{\dagger}_{j,j_z}$) are the
electron annihilation and creation operators at orbital $|P_{j}, j_z\rangle$ ($|D_{j}, j_z\rangle$). The diagonal term of $P$ orbitals $E_{jj_z}^{p}=\frac{\lambda_p}{3}$ if $j=\frac{3}{2}$, and $E_{jj_z}^{p}=-\frac{2\lambda_p}{3}$ if $j=\frac{1}{2}$. The diagonal term of $t_{2g}$ orbitals $E_{jj_z}^{d}=E_d-\frac{\lambda_d}{3}$ if $j=\frac{3}{2}$, and $E_{jj_z}^{d}=E_d+\frac{2\lambda_d}{3}$ if $j=\frac{1}{2}$. The parameter $\lambda_{p}$ and $\lambda_{d}$ is the SOC parameters for N $2p$ and Tl $5d$, respectively. $E_d$ is the on-site energy
of $5d$ orbitals and that of $p$ is taken as zero.

%\begin{eqnarray*}
%  H_{pd}&=  \left(\begin{array}{cc}
%      H_{pd}^{+}        &0       \\
%      0        &H_{pd}^{-}       \\
%\end{array}\right),
%\end{eqnarray*}
%where $H_{pd}^{\pm}$ is in basis with positive and negative $j_z$ quantum number, respectively. Both of them are in %the form of
%\begin{eqnarray*}
%  H_{pd}^{\pm}&=  \left(\begin{array}{cccccc}
%      \frac{\lambda_p}{3}        & t_{pd}    &0    &0    &0    &0     \\
%      t_{pd}        & E_d-\frac{\lambda_d}{3}    &0    &0    &0    &0     \\
%      0        &0    &\frac{\lambda_p}{3}        & t_{pd}    &0    &0     \\
%      0        &0   & t_{pd}        &E_d-\frac{\lambda_d}{3}    &0    &0     \\
%      0        &0    &0    &0    &-\frac{2}{3}\lambda_p    & t_{pd}     \\
%      0        &0    &0    &0    & t_{pd}    &E_d+\frac{2}{3}\lambda_d     \\
%\end{array}\right),
%\end{eqnarray*}
%where $\lambda_{p}$ and $\lambda_{d}$ is the SOC parameters for N $2p$ and Tl $5d$, respectively. $E_d$ is the on-site %energy
%of $5d$ orbitals and that of $p$ is taken as zero.

For the Hamiltonian, Tl-5$d$ SOC could be obtained by FLAPW calculation as $\lambda_d$=2.168 eV, and $E_d$ takes the value of -10.0 eV, which is approximately the gravity center of Tl $5d$-orbitals in Fig.\ref{band} (c) and (d). Depending on the strength of $p$-$d$ hybridization $t_{pd}$ and $\lambda_p$, the system falls into two different phases, as shown in Fig.\ref{phase}(a). In the strong $p$-$d$ hybridization limit, the effective SOC $\lambda_{eff}$ of anion's $p$ orbitals is negative and dominated by the
 SOC of cation's 5$d$ orbitals. The $p$ states split into a $j_{eff}$=1/2 doublet and a $j_{eff}$=3/2
quartet states with the former energetically higher. The band structure opens a gap in the whole BZ, and the system is a TI.  In the strong $\lambda_p$ limit, $\lambda_{eff}$ is dominated by the
atomic SOC of $p$ orbitals, and it is positive.  The $j_{eff}$=1/2 doublet states is energetically
lower than the $j_{eff}$=3/2 quartet states, leading the system in a topological semimetal rather than a true insulator.  Between the two regions, there should exist a critical line with $\lambda_{eff}$=0. In other words, $t_{pd}$ and $\lambda_p$ are in balance. For material realization, TlN and HgS are at TI region, TlAs and HgTe are at TSM region, and TlP is almost at the borderline.

%We have obtained $\lambda_p$=0.018 eV and $\lambda_d$=2.168 eV by FLAPW calculation for single N and Tl atom, %respectively.
%$E_d$ takes the value of -10.0 eV, which is approximately the gravity center of Tl $5d$-orbitals in Fig.\ref{band}. %The effective SOC
%$\lambda_{eff}$ of N $2p$ orbitals due to $p$-$d$ hybridization strongly depends on the strength of hybridization
%$t_{pd}$ as shown in Fig.\ref{phase}(a). We find that if $|t_{pd}|<0.88$ eV, $\lambda_{eff}$ is positive and dominated %by the
%atomic SOC of $p$ orbitals. Once the hybridization is stronger than 0.88 eV, $\lambda_{eff}$ becomes negative.
%From band structure of first-principles calculations, the estimated $\lambda_{eff}$ is -0.036 eV and the corresponding
%$t_{pd}$ is about 1.57 eV, which is consistent with the band width, $\sim$2.5 eV, of Tl $5d$ orbitals.

%We can also study the dependence of $\lambda_{eff}$ on the atomic SOC strength $\lambda_{p}$ of anion by fixing
%$t_{pd}$=1.57 eV and other parameters the same as above. As shown in Fig.\ref{phase}(b), $\lambda_{eff}$ increases %almost
%linearly with $\lambda_{p}$ from negative to positive. The critical value of $\lambda_{p}$ is 0.055 eV, where
%$\lambda_{eff}$=0.0 eV. As we noticed that the atomic SOC of phosphorus is 0.052 eV, very close to the critical value.
%We have calculated Wurtzite TlP and TlAs. The former has nearly zero effective SOC in P $3p$ orbitals and the later %one
%has positive value. Both of them are metal with $s$-$p$ band inversion. (see Supplemental Material for details)

\textit{Topological phase transition by tuning effective SOC}.
As SOC plays the critical role in determining the topology of bands, tuning its strength and sign can introduce nontrivial
topological phase transition. To show how this can happen in realistic material TlN, an effective {\it ab initio}
tight-binding Hamiltonian base on MLWFs of Tl-$6s$ and N-$2p$ orbitals has
been established. The effective SOC $\lambda_{eff}$ is then added onto N-$2p$ MLWFs as a tunable parameter. As shown
in Fig.\ref{phase}(b), taking $\lambda_{eff}$=-36 meV can reproduce the GGA+SOC calculation very well and TlN is a TI.
When $\lambda_{eff}$ is zero, this reproduces the GGA calculation, the band inversion and the double degeneracy of $p_x$
and $p_y$ orbitals lead to zero gap semimetal state, as shown in Fig.\ref{phase}(c). When $\lambda_{eff}$ is taken as 36 meV,
it becomes a topological 3D Dirac semimetal with Dirac point on the path $\Gamma$-A. The sign reversal of $\lambda_{eff}$
leads to the energy order reversal of $\Gamma_{9}^{\frac{3}{2}}$ and $\Gamma_{7}^{\frac{1}{2}}$ states.
When $\Gamma_7^{\frac{1}{2}}$ is higher ($\lambda_{eff}<$0), the inverted $\Gamma_7^{s}$ can have hybridization
with it in all direction of BZ and it is TI. When $\Gamma_9^{\frac{3}{2}}$ is higher ($\lambda_{eff}>$0), $\Gamma_7^{s}$
bands exactly cross $\Gamma_9^{\frac{3}{2}}$ at the Dirac point since they belong to $\Delta_7$
and $\Delta_9$ irreducible representation, respectively, which are distinguished by the C$_3$ rotation symmetry along $\Gamma$-A.
Such band inversion resulted 3D Dirac semimetal state protected by crystal symmetry is the similar as those in Na$_3$Bi
and Cd$_3$As$_2$.~\cite{Na3Bi, Cd3As2} We have found that when lattice strain is of
$a$=1.06$a_0$ and $c$=1.10$c_0$ ($a_0$ and $c_0$ are experimental lattice constants), the band structure from
first-principles calculation is the same as that in Fig.\ref{phase}(d) and TlN becomes a 3D Dirac semimetal
(see Supplemental Material for details). 3D Dirac semimetal is a symmetry-protected topological state with a single pair of 3D Dirac points in the bulk and non-trivial Fermi arcs on the surfaces. The 3D Dirac point can be described as four-component Dirac fermions, which can be viewed as two copies of distingct Weyl fermions. Therefore, Weyl semimetal could be realized based on 3D Dirac semimetal when time-reversal or inversion symmetry was broken.~\cite{Na3Bi, Cd3As2,weylsemimetal} It can also be driven into TI by symmetry breaking, and quantum spin Hall effect in its quantum well structure.
%It should be noted that such lattice strain doesn't break any crystal symmetry while it is impossible in Zinc-blende structure.

In summary, we have shown that TlN is a 3D TI with effectively ``negative" SOC. The band inversion 
between Tl-$6s$ and N-$2p$ orbital is robust and the $p$-$d$ hybridization between
N $2p$ and Tl $5d$ determines its effective SOC. Its surface state has {\it right-handed}
helical spin-momentum locking, which is opposite to other known TIs. Heterostructure formed by two TIs 
with opposite spin-momentum locking helicity might host novel phenomena. The effective SOC can be 
tuned to be positive by suitable lattice strain without breaking any crystal symmetry and drives TlN from 
TI to 3D Dirac semimetal. These make TlN quite unique and a good playground for further study.

The author XLS acknowledges helpful discussions with L. R. Shaginyan, Fei Ye and Jian-Zhou Zhao. HMW acknowledges the
hospitality during his stay in NICHe, Tohoku University. We acknowledge the supports from the NSF of China and the 973
program of China (No. 2011CBA00108 and 2013CB921700).

%\textit{Note}. After we finished the preparing of this manuscript, we became aware that a very recent work~\cite{wangHgTe} of Wang {\it et al.}
%discussed the opposite spin-moment locking effect in HgTe and HgS and have studied the gapless interface states
%of them protected by different mirror Chern number.

\newpage

\section{Supplemental Material}

\subsection{$\mathbb{Z}_2$ Indices by Wilson Loop Method}

\begin{figure}[tbp]
\includegraphics[clip,scale=0.45,angle=270]{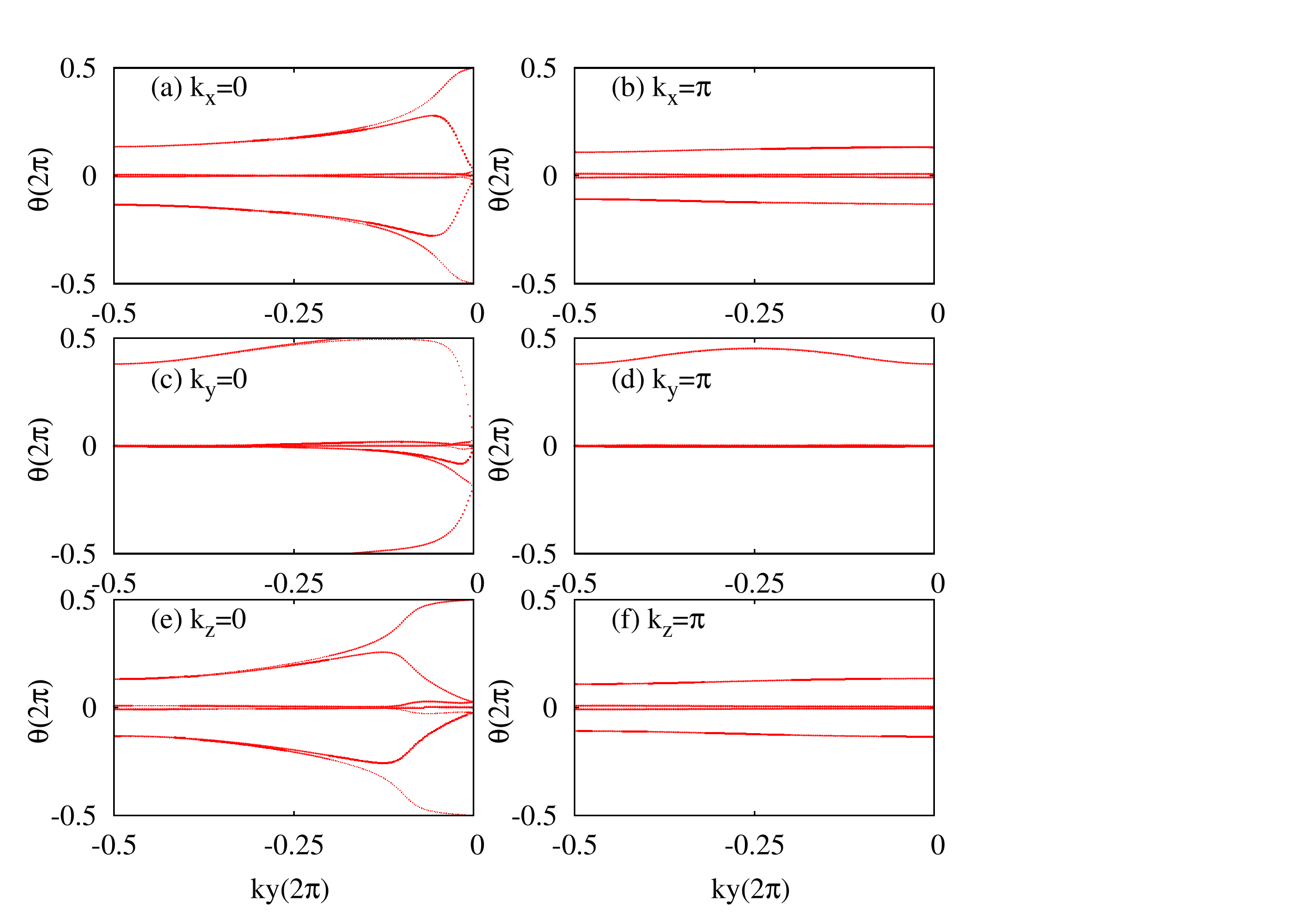}
%\centerline{\includegraphics[width=8.0cm,angle=270]{fig4.eps}}
\caption{(Color online) The Wannier function center evolution on BZ planes (a) $k_x=0$, (b) $k_x=\pi$, (c) $k_y=0$, (d) $k_y=\pi$, (e) $k_z=0$, and (f) $k_z=\pi$. It demonstrates the $\mathbb{Z}_2$ indices as (1;000).}
\label{fig:z2}
\end{figure}

Band inversion is a necessary condition for the existence of topological nontrivial states, but it is not sufficient to distinguish a topological insulator (TI), because topological invariant is a global character of the electronic structure in the whole Brillouin zone (BZ).  The definition of parity criterion is convenient to identify TIs, but it can not be used in Wurtzite structure (or Zinc-blend) because of the lack of inversion symmetry. The Wilson loop method can be employed to study the evolution of Wannier function center~\cite{YuruiZ2, tireviewofus}. We calculated the Wannier center evolution for the six BZ plane of TlN, as shown in Fig.\ref{fig:z2}, and obtain $\mathbb{Z}_2$ indices as (1;000), demonstrating that TlN is a 3D strong TI.

\subsection{First-principles Energy Bands with lattice strain of $a$=1.06$a_0$ and $c$=1.10$c_0$. }

\begin{figure}[tbp]
\includegraphics[clip,scale=0.32,angle=0]{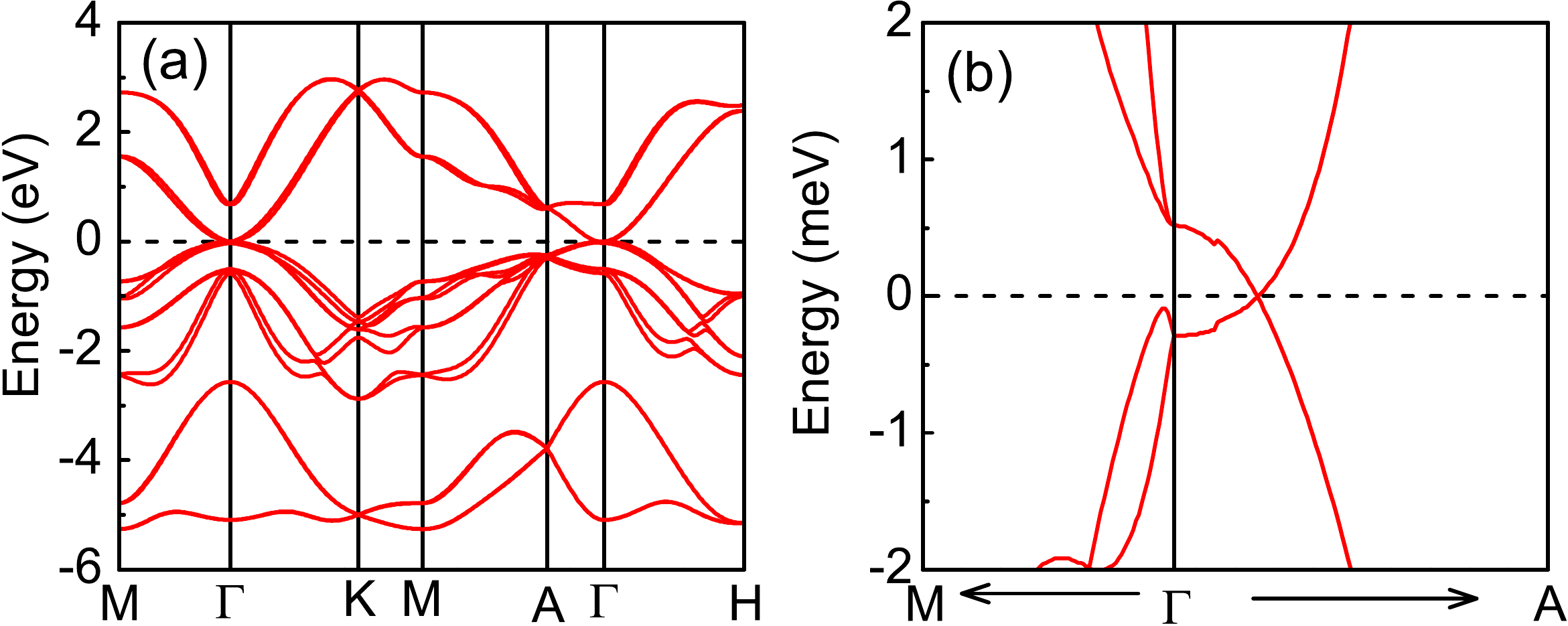}
%\centerline{\includegraphics[width=8.0cm,angle=270]{fig4.eps}}
\caption{(Color online) (a) The band structure of strained TlN with $a$=1.06$a_0$ and $c$=1.10$c_0$. (b)Partially enlarged view of (a).}
  \label{fig:TlNac}
\end{figure}

To tune the effective SOC parameter of TlN, we can enlarge the lattice constants to reduce the $p$-$d$ hybridization. From realistic first-principles calculation, we found that if $a$=$a_0$ and $c\geq$1.150$c_0$ (or $c$=1.10$c_0$ and 1.04$a_0$$\leq$$a$$\leq$1.06$a_0$) TlN would be in 3D Dirac semimetal state. The band structure of TlN with $a$=1.06$a_0$ and $c$=1.10$c_0$ is shown in Fig. \ref{fig:TlNac}. It is clear that there is a 3D Dirac point on $\Gamma$-$A$ path.

\subsection{Energy Bands of TlP and TlAs}
\begin{figure}[tbp]
\includegraphics[clip,scale=0.32,angle=0]{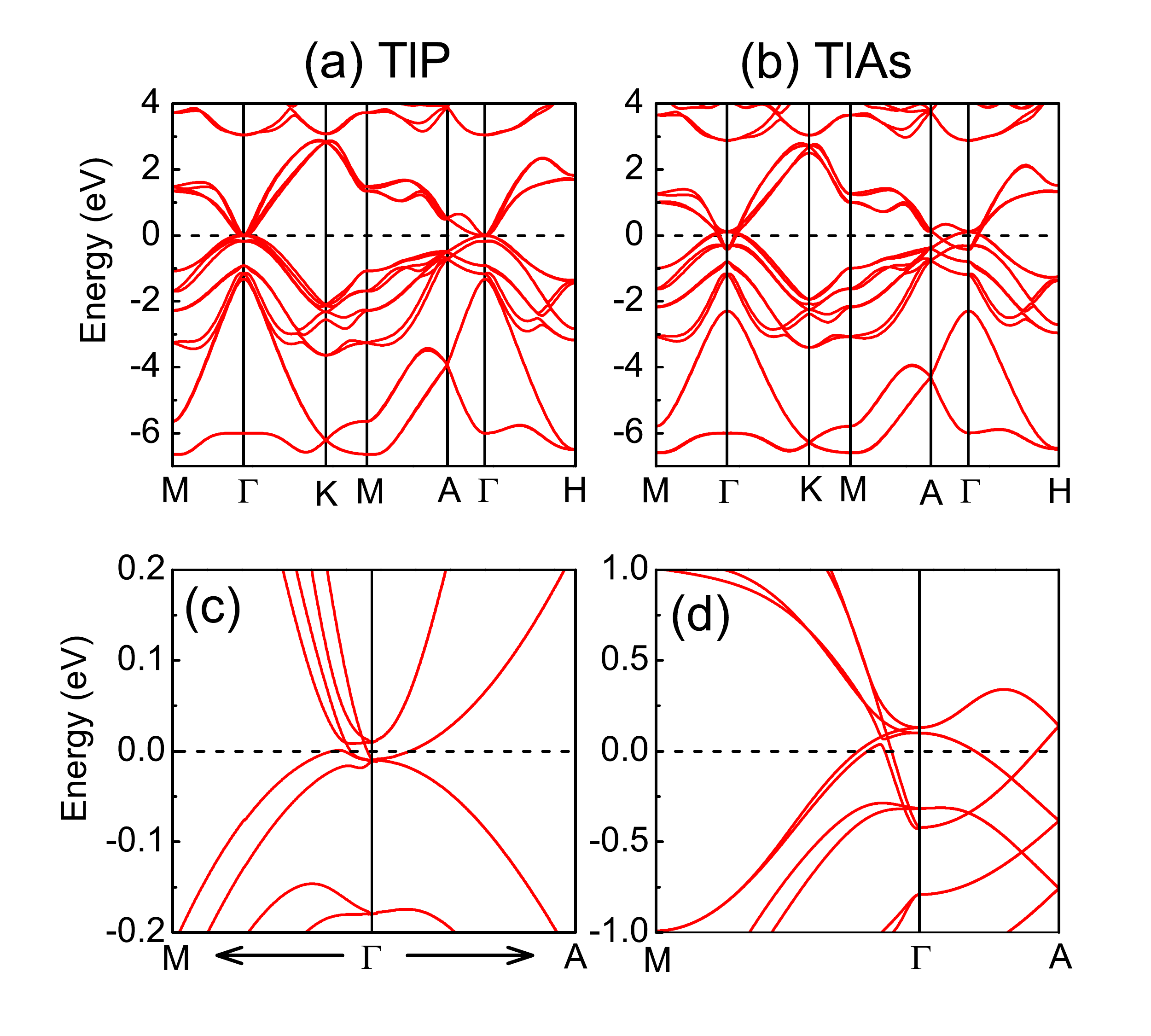}
%\centerline{\includegraphics[width=8.0cm,angle=270]{fig4.eps}}
\caption{(Color online) The band structure of (a) TlP and (b) TlAs. (c) and (d) are the partially enlarged view of (a) and (b), respectively.}
  \label{fig:TlPTlAs}
\end{figure}

As TlN is a TI, it is natural to ask how about TlP and TlAs. We have found that the total energy of Wurtzite structure is lower than that of Zinc-blende structure for both of TlP and TlAs with relaxed structure. Band inversion occurs in both of them, but there is no finite band gap as shown in Fig.\ref{fig:TlPTlAs}. The effective SOC for TlP is nearly zero, because the intrinsic SOC of P-$3p$ orbital is comparable to the $p$-$d$ hybridization, which is in good agreement with model analysis in main text.

\subsection{Tight-binding Energy Bands by tuning effective SOC}

\begin{figure}[tbp]
\includegraphics[clip,scale=0.32,angle=0]{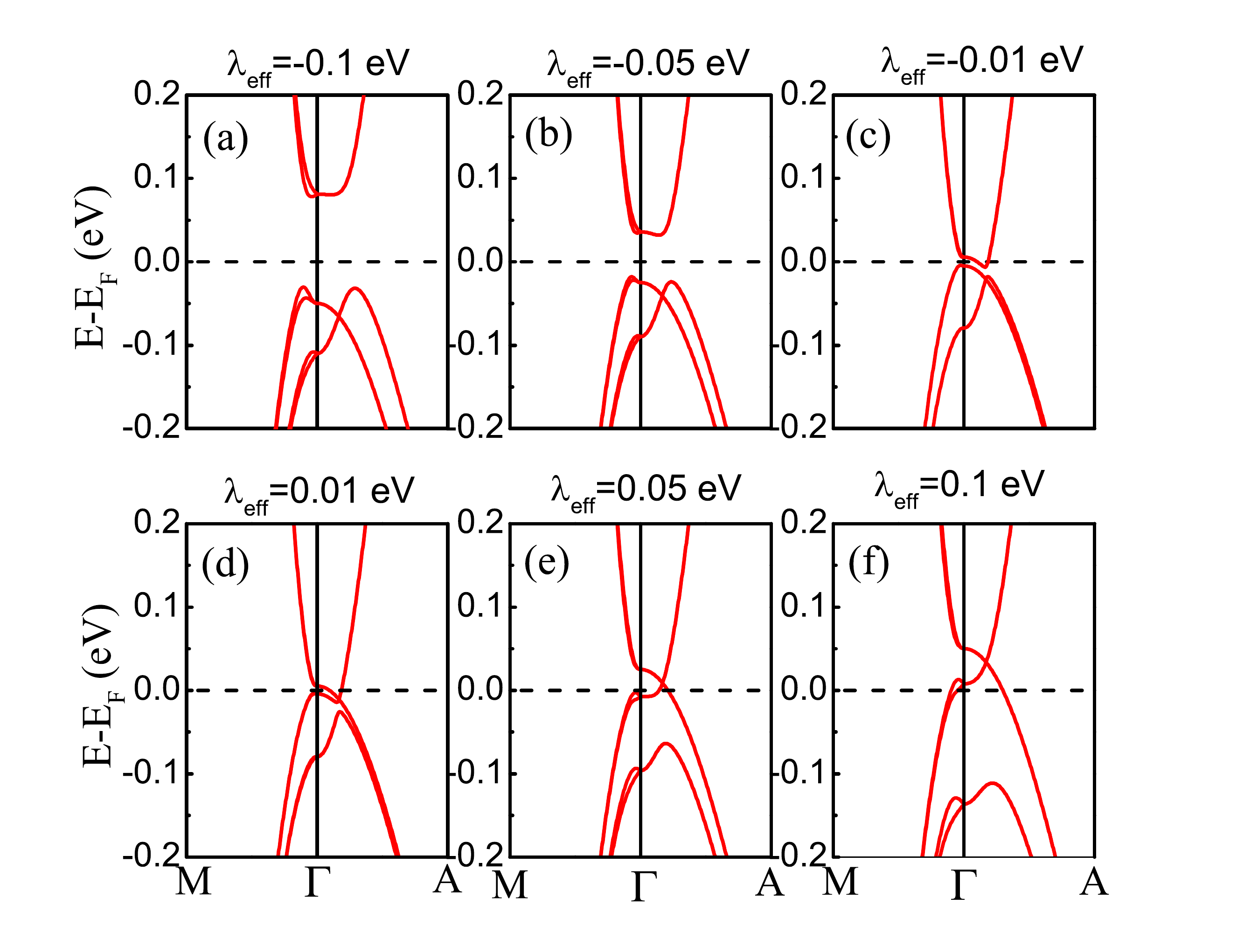}
%\centerline{\includegraphics[width=8.0cm,angle=270]{fig4.eps}}
\caption{(Color online) The band structure of Wannier tight-binding model with effective SOC parameter as (a) $\lambda_{eff}$=-0.1 eV, (b) $\lambda_{eff}$=-0.05 eV, (c) $\lambda_{eff}$=-0.01 eV, (d) $\lambda_{eff}$=0.01 eV, (e) $\lambda_{eff}$=0.05 eV and (f) $\lambda_{eff}$=0.1 eV.}
  \label{fig:wannier}
\end{figure}

The phase diagram in Fig. 4 in main text is obtained by using the tight-binding model based on Wannnier functions with different
effective SOC parameters. As shown in Fig.\ref{fig:wannier}, it is a 3D Dirac semimetal when $\lambda_{eff} > 0$, otherwise it is a TI.

\subsection{Kane Model}

The 8$\times$8 Kane model can be used to describe the band structure of TlN. Firstly, a 4$\times$4 model without SOC can be constructed in the basis set as $|S\rangle$, $|P_{+}\rangle$, $|P_{-}\rangle$, $|P_{z}\rangle$, where $|P_{\pm}\rangle=\frac{1}{\sqrt{2}}(|P_x\rangle \pm i|P_y\rangle)$. The Hamiltonian reads

\begin{eqnarray*}
  H_{4\times4}({\vec{k}})&=  \left(\begin{array}{cccc}
      E_s       & ip_1k_{+}         & ip_1k_{-}         & d+ik_zp_2 \\
      -ip_1k_{-}& \lambda+\theta    &a_5k_{-}^2         & a_6k_zk_{-}  \\
      -ip_1k_{+}& a_5k_{+}^2        & \lambda+\theta    & a_6k_zk_{+}\\
      d-ip_2k_z & a_6k_zk_{+}       & a_6k_zk_{-}       & \lambda-\delta
\end{array}\right)
\end{eqnarray*}

The 8$\times$8 model with SOC can be constructed as
$H_{8\times8}({\vec{k}})=I\bigotimes H_{4\times4}({\vec{k}})+H_{so}$.  $H_{so}$ has the following form in the basis order: $|S_\uparrow\rangle$, $|P_{+\uparrow}\rangle$, $|P_{-\uparrow}\rangle$, $|P_{z\uparrow}\rangle$, $|S_\downarrow\rangle$, $|P_{+\downarrow}\rangle$, $|P_{-\downarrow}\rangle$, $|P_{z\downarrow}\rangle$.

\begin{eqnarray*}
  H_{so}&=  \frac{\xi}{2}\left(\begin{array}{cccccccc}
      0    &0    &0    &0    &0    &0    &0    &0     \\
      0    &1    &0    &0    &0    &0    &0    &0     \\
      0    &0    &-1    &0    &0    &0    &0    &\sqrt{2}     \\
      0    &0    &0    &0    &0    &-\sqrt{2}    &0    &0     \\
      0    &0    &0    &0    &0    &0    &0    &0     \\
      0    &0    &-\sqrt{2}    &0    &0    &-1    &0    &0     \\
      0    &0    &0    &0    &0    &0    &1    &0     \\
      0    &\sqrt{2}    &0    &0    &0    &0    &0    &0
\end{array}\right)
\end{eqnarray*}

where $E_s=E_s+s_1k_z^2+s_2k_{+}^2$, $\lambda=E_p+a_1k_z^2+a_2k_{+}^2$, $\theta=a_3k_z^2+a_4k_{+}^2$. The parameters take values from fitting first-principles results as $E_s$=-2.05842, $E_p$=0.00015, $\delta$=0.07903, $s_1$=8.0, $s_2$=3.0, $a_1$=-2.0, $a_2$=-5.3344, $a_3$=0.0, $a_4$=0.1, $a_5$=0.3, $a_6$=0.2, $p_1$=2.9698, $p_2$=6.3, $\xi$=-0.025. It is noticed that the effective SOC parameter $\xi$ is negative.

The atomic Tl-$6s$ and N-$2p$ states with SOC can be written as the
states with definite angular momentum $J$ and $J_z$, i. e.,
$|S_{J=\frac{1}{2}},J_z=\pm\frac{1}{2}\rangle$,
$|P_{\frac{3}{2}},\pm\frac{3}{2}\rangle$,
$|P_{\frac{3}{2}},\pm\frac{1}{2}\rangle$,
$|P_{\frac{1}{2}},\pm\frac{1}{2}\rangle$.
In Wurtzite TlN, the valence band is mainly contributed by $|P_{\frac{1}{2}},\pm\frac{1}{2}\rangle$, and inverted states is $|S_{\frac{1}{2}},\pm\frac{1}{2}\rangle$. Therefore, the inversion mechanism can be described by
the  $|P_{\frac{1}{2}},\pm\frac{1}{2}\rangle$ and the $|S_{\frac{1}{2}},\pm\frac{1}{2}\rangle$ states. We can construct a low energy Hamiltonian around $\Gamma$ point, by
 considering only the minimal basis set of
$|S_\frac{1}{2},\frac{1}{2}\rangle$,
$|P_{\frac{1}{2}},\frac{1}{2}\rangle$,
$|S_\frac{1}{2},-\frac{1}{2}\rangle$ and
$|P_{\frac{1}{2}},-\frac{1}{2}\rangle$ states.   We can directly get it by downfolding the 8-band model into
the subspace spanned by the 4 mimimal basis.  The resulting
$H_\Gamma({\vec k})$ reads,
%in set $\mathcal{A}$
\begin{eqnarray*}
\label{H44}
  H_{\Gamma}({\vec{k}})&=&\epsilon_0(\vec{k}) +\left(\begin{array}{cccc}
      M(\vec{k}) & iAk_{z} & Dk_{-} & -iAk_{-} \\
      -iAk_{z} & -M(\vec{k}) & -iAk_{-}  & Dk_{-} \\
      Dk_{+} & iAk_{+}  & M(\vec{k}) & iAk_{z}\\
      iAk_{+} & Dk_{+} & -iAk_{z} & -M(\vec{k})
\end{array}\right)
\end{eqnarray*}
where $k_{\pm}=k_{x}\pm ik_{y}$ and
$M(\vec{k})=M_{0}-M_{1}k_z^2-M_{2}(k_{x}^{2}+k_y^{2})$ with parameters
$M_0$, $M_1$, $M_2$ $<$0 to reproduce band inversion. The parameter D is induced to describe the breaking of inversion symmetry.  In such
case, the energy dispersion is
$E(\vec{k})=\epsilon_0(\vec{k})\pm\sqrt{M(\vec{k})^{2}+A^{2}\vec{k}^2+D^2k_{+}k_{-}+2\sqrt{D^2(M^2+A^2k_{z}^2)k_{+}k_{-}}}$.

We can take the values as $M_0=-0.1$, $M_1=-10$, $M_2=-10$, $A=0.5$ and $D=0.1$, to study the low energy physics qualitatively.
\end{document}